\documentclass{article}
\usepackage[latin1]{inputenc}
\usepackage[english]{babel}
\usepackage{dsfont}
\usepackage[cyr]{aeguill}
\usepackage{amsmath,amssymb}
\usepackage{amsthm}
\usepackage{eurosym}
\usepackage{multicol}
\usepackage{color}
\usepackage{fancyhdr}
\usepackage[pdftex]{graphicx}
\usepackage[pdftex]{hyperref}
\usepackage{verbatim}
\usepackage{setspace}
\usepackage{stmaryrd}

\newcommand{\indic}{\mathds{1}} 

\theoremstyle{plain}

\providecommand{\keywords}[1]{\textbf{\textit{Keywords --}} #1}

\title{Estimation of time-varying kernel densities \\ and chronology of the impact \\ of COVID-19 on financial markets}

\author{Matthieu Garcin$^{\text{a,}}$\thanks{ Corresponding author: matthieu.garcin@m4x.org. \newline $^{\text{a}}$ Léonard de Vinci Pôle Universitaire, Research center, 92916 Paris La Défense, France.\newline $^{\text{b}}$ ESILV, 92916 Paris La Défense, France. \newline The authors would like to thank Brieuc-Marie Le Brigand for his valuable help in the implementation of some of the methods described in this paper.}, Jules Klein$^{\text{b}}$, Sana Laaribi$^{\text{b}}$}

\date{\today}

\begin{document}

\maketitle

\begin{abstract}
The time-varying kernel density estimation relies on two free parameters: the bandwidth and the discount factor. We propose to select these parameters so as to minimize a criterion consistent with the traditional requirements of the validation of a probability density forecast. These requirements are both the uniformity and the independence of the so-called probability integral transforms, which are the forecast time-varying cumulated distributions applied to the observations. We thus build a new numerical criterion incorporating both the uniformity and independence properties by the mean of an adapted Kolmogorov-Smirnov statistic. We apply this method to financial markets during the onset of the COVID-19 crisis. We determine the time-varying density of daily price returns of several stock indices and, using various divergence statistics, we are able to describe the chronology of the crisis as well as regional disparities. For instance, we observe a more limited impact of COVID-19 on financial markets in China, a strong impact in the US, and a slow recovery in Europe.
\end{abstract}


\section{Introduction}

The knowledge of the distribution of price returns is overriding in finance. Indeed, forecasts and risk measures, such as the Value-at-Risk (VaR), the expected shortfall, or even the volatility, can be seen as scalars calculated from a probability density function (pdf). Practitioners appreciate these scalars for their simplicity but the pdf contains relevant and more comprehensive information. An accurate description of the pdf is thus worthwhile in a financial perspective.

For this reason, pdf in finance should not be limited to the popular Gaussian distribution. More realistic parametric distributions have thus been put forward~\cite{MSK}, like the NIG distribution~\cite{FB} or the alpha-stable one~\cite{TRS}, among many others. Besides, the non-parametric alternative makes it possible to depict more accurately the real pdf but it may be subject to overfitting if it does not include any regularization. For this purpose, Beran has proposed the minimum Hellinger distance approach, in which a non-parametric pdf has to be estimated first, before being approximated by a parametric distribution~\cite{Beran}. This approach finds some applications in finance~\cite{LB}. Other semi-parametric approaches include the distortion of a parametric density in order to take into account higher-order empirical moments, using for example an Edgeworth expansion~\cite{LAC,GG14}, which also has some applications in finance~\cite{NP}. Finally, non-parametric approaches, like the kernel density, include a smoothing parameter, called bandwidth, which is supposed to balance accuracy and statistical robustness~\cite{WJ,Tsybakov}. Selecting an appropriate bandwidth is a hard task often left aside by practitioners in finance, but some statistical methods propose criteria that the bandwidth should minimize, like the asymptotic mean integrated square error (AMISE)~\cite{JMS}.

In finance, non-parametric methods are not limited to the estimation of a pdf. We can cite for example their use to estimate the impact of market events, as with the non-parametric news impact curve in econometric volatility models~\cite{PS,FY,BM,3V}. The rationale of such an approach is that linear impact models misestimate the reaction of markets to extreme events. Similarly, parametric models do often not describe accurately enough the tails of the pdf of price returns. Extreme events may also lead to other methodological choices in addition to the non-parametric approach. Indeed, like all the previous financial crisis, the recent crisis provoked by the COVID-19 pandemic has highlighted that the occurrence of an extreme event may have a sustainable effect on the market, namely on the pdf of price returns. Bad economic news do not only result in one extreme daily price variation but it can initiate a longer turmoil period. This alternation of market regimes incites to introduce time-varying densities. Once again, several approaches are possible, depending on the fact that we consider a parametric pdf~\cite{AABBG} or a non-parametric one~\cite{HO}. We are particularly interested in the non-parametric approach, which offers the possibility to reach a higher accuracy. We stress the fact that applications of time-varying kernels in finance are not limited to estimating a pdf of price returns. They indeed include interest rate models~\cite{ZW} or the study of the dynamic pdf of correlation coefficients with a particular shape during tumble periods~\cite{LLT}.

Time-varying kernel densities rely on the choice of two important free parameters: the bandwidth smooths the pdf at a given date, exactly as in the static approach, whereas the discount factor smooths the variations in time. Harvey and Oryshchenko propose a maximum likelihood approach to select these free parameters~\cite{HO}. However, the literature about density validation requires stronger properties, namely the fact that the cumulative distribution function (cdf) of price returns must form a set of iid uniform variables, known as the probability integral transforms (PITs)~\cite{DGT}. In the present paper, we thus propose a new selection rule of the bandwidth and of the discount parameter, so that it is consistent with the validation rule of the pdf. The main challenge is then to build the criterion, that is to define a function of the bandwidth and of the discount factor that we intend to minimize. Indeed, the traditional approach for validating an estimated pdf consists in a series of statistical tests and graphical analysis, not in a sole numerical criterion. The criterion we propose relies on a Kolmogorov-Smirnov statistic, that we can replace by any statistic of distribution divergence. We also adapt this statistic so as to minimize the discrepancy of the series of the PITs for which we need the independence. Our work is not the first one to stress the limitations of the maximum likelihood approach in the selection of the two free parameters. We can indeed cite an article following a least-square approach~\cite{SO} and another one maximizing a uniformity criterion for the PITs by the mean of an artificial neural network~\cite{WTS}. Our method differs from the latter as it also takes into account the independence of the PITs and it requires only standard statistical tools compared to artificial neural networks.

We apply our new method to several stock indices before and during the financial crisis induced by the onset of the COVID-19 in the US, in Europe, and in Asia. The question of the impact of the pandemic on stock markets is a hot topic. Several papers deal with this subject and stress the exceptional amplitude of the crisis~\cite{AANH,BBDK,Pav}. We propose here a new outlook on this financial crisis, using the time-varying kernel densities to describe its chronology. We also study the significance of the daily kernel density with respect to the pdf in a steady market. This makes it possible to determine the interval of dates for which the distribution of price returns significantly indicates a financial crisis. In particular, we observe that the speed at which markets recover varies a lot among the regions considered.

The paper is organized as follows. In Section~\ref{sec:Stat}, we introduce the method for estimating a dynamic kernel density along with the selection rule for the bandwidth and the discount factor. In Section~\ref{sec:simul}, we compare this selection method to another one based on a likelihood criterion, with the help of simulations. In Section~\ref{sec:EmpRes}, we apply the introduced method to stock markets during the COVID-19 crisis. Section~\ref{sec:concl} concludes.

\section{Statistical methods}\label{sec:Stat}

In this section, we introduce a method to estimate a time-varying density. For this purpose, we recall how we can estimate a static non-parametric density as well as its dynamic adaptation. This method relies on the choice of two free parameters. The main innovation of this paper consists in basing this choice on a quantitative version of criteria usually devoted to the evaluation of forecast densities. The last subsection is about some divergence statistics between two densities. We will use these divergences in the empirical part of this paper to quantify the amplitude of the variations of the densities through time and to determine the significance of these variations.

\subsection{Kernel density}

A widespread non-parametric method to estimate a pdf uses kernels. The kernel density, estimated on observations $X_1,...,X_t$, is defined, for $x\in\mathbb R$, by:
\begin{equation}\label{eq:KernelStatique}
\hat f(x)=\frac{1}{th}\sum_{i=1}^{t}{K\left(\frac{x-X_i}{h}\right)},
\end{equation}
where $h>0$ is the bandwidth and $K$ a function following the same rules as a pdf, namely it is positive, integrable and its integral is one~\cite{WJ,Tsybakov}. With these two properties, $\hat f$ also has the features of a density. In particular, when integrating $\hat f$, the substitution $y=(x-X_i)/h$ in each of the $t$ integrals in the sum clearly shows that we need to normalize the sum by $th$ in order to have $\int_{\mathbb R}{\hat f(x)dx}=1$. The symmetry and the continuity of the kernel is also often desirable.

The rationale of the kernel density is to make a continuous generalization of a histogram. Indeed, in the histogram, we count the number of occurrences in given intervals. The thinner the intervals, the more accurate the density estimation. But very thin intervals lead to overfitting, with a very erratic estimated density. To avoid this, we prefer to smooth the histogram. A simple manner to do this consists in replacing the number of occurrences of observations in each thin interval by a criterion of proximity of each observation to the middle of this interval. This is how the kernel density works. The \textit{proximity} function $K$ must thus reach its maximum in zero and it must decrease progressively when its argument gets away from zero. Thus, the impact of $X_t$ on the estimated pdf in $x$ is maximal for $x=X_t$ and it decreases progressively when $|x-X_t|$ becomes higher until reaching zero, at least asymptotically. It means that the observation of $X_t$ will have no impact on the density $\hat f(x)$ if $X_t$ is by far greater or lower than $x$.

There exists a large literature on the choice of the kernel $K$~\cite{Scaillet,BR}. Epanechnikov and Gaussian kernels are widespread, due to their simplicity.\footnote{ In the empirical application, we use Epanechnikov kernel.} But it seems, according to the related literature, that the choice of the kernel is often less overriding than the choice of the bandwidth $h$. Indeed, this parameter plays the role of a regularization parameter. In practice, we tune $h$ in order to balance accuracy and robustness. The larger $h$, the wider each kernel and the larger the interval on which each observation has an impact. We review in Section~\ref{sec:bandwidth} some methods to select this bandwidth.

\subsection{Dynamic kernel density}

We can change the formulation of the kernel density in order to take into account its progressive evolution through time. We get this dynamic version of the kernel density by the mean of weights $w_{t,i}$: 
\begin{equation}\label{eq:InitDens}
\hat f_t(x)=\frac{1}{h}\sum_{i=1}^{t}{w_{t,i} K\left(\frac{x-X_i}{h}\right)},
\end{equation}
such that $\sum_{i=1}^{t}{w_{t,i}}=1$~\cite{HO}. For a fixed $t$, if the weights $w_{t,i}$ increase with $i$, more recent observations will be overweighted and the update of the kernel density is consistent with the economic intuition. The exponential weighting is widespread in the statistical literature as it can reduce the computation of a density update to a simple recursive formula instead of the linear cost induced by the whole estimation of the density from scratch. We then express the weights by 
$$w_{t,i}=\frac{1-\omega}{1-\omega^t}\omega^{t-i},$$ 
where $0<\omega\leq 1$. With this setting for the weights, we note $\hat f^{h,\omega}_t$ the density introduced in equation~\eqref{eq:InitDens}. Then, when the duration of the estimation sample is large enough with respect to the speed of decay of the weights, a good approximation of $w_{t,i}$ is $(1-\omega)\omega^{t-i}$. The recursive formula that the dynamic kernel density follows is then:
\begin{equation}\label{eq:MajDens}
\hat f^{h,\omega}_{t+1}(x)=\omega \hat f^{h,\omega}_t(x)+\frac{1-\omega}{h}K\left(\frac{x-X_{t+1}}{h}\right).
\end{equation}
The two free parameters of this dynamic non-parametric density are the bandwidth $h$, and the discount factor $\omega$. Starting at a given time $t_0$ from a density with an exponential weighting, such as in equation~\eqref{eq:InitDens}, we obtain the density at subsequent times iteratively by applying equation~\eqref{eq:MajDens}.

Along with the time-varying density, we can build the corresponding cdf. Integrating equation~\eqref{eq:InitDens}, the first estimated cdf, at time $t_0$, is:
$$\hat F^{h,\omega}_{t_0}(x)=\frac{1-\omega}{1-\omega^{t_0}}\sum_{i=1}^{t_0}{\omega^{t_0-i} \mathcal{K}\left(\frac{x-X_i}{h}\right)},$$
where $\mathcal K$ is the primitive of $K$ such that $\underset{x\rightarrow \infty}{\lim}\mathcal K(x)=1$. Subsequently, we get the cdf at a time $t+1$ by the mean of the following iteration:
$$\hat F^{h,\omega}_{t+1}(x)=\omega \hat F^{h,\omega}_t(x)+(1-\omega)\mathcal K\left(\frac{x-X_{t+1}}{h}\right),$$
which is the primitive of equation~\eqref{eq:MajDens}.

Other approaches are possible for estimating a time-varying density. For example, we could have estimated static densities on successive intervals and have then smoothed the transition between the resulting densities. For parametric densities, this amounts to smoothing time-varying parameters, which is a well-known subject in statistics~\cite{Garcin2017}. However, this approach is not very natural for non-parametric densities, as we need a big amount of data to estimate one static density.

\subsection{Evaluation of the quality of the dynamic density}

The purpose of density forecast may vary a lot. In a financial perspective, one may need it to build risk measures, or to forecast an average price return or a most likely price return. In practice, an investment decision is to be made relying on this density. One must then evaluate the quality of the forecast with respect to a loss function corresponding to this decision. Unfortunately, there cannot exist any absolute ranking of density forecasts valid for all the possible loss functions~\cite{DGT}. 

We therefore have to make a choice which is necessarily subject to discussion. In the econometric literature, we can find an evaluation of the density forecast by the mean of the likelihood of observations~\cite{HO}. We think that this choice, as focusing on the body of the distribution, neglects the behaviour of the density in its tails. A more general perspective would incite to choose a density forecast consistent with the real density, even with its tails. Such a forecast would be more relevant in finance for calculating a VaR or an expected shortfall. However, the real density is never observed. If we had a static density forecast, the evaluation of this forecast could simply consist in comparing it with the empirical density of all the observed price returns. But the forecast density is supposed to change at each time, so that we have to base our analysis on another invariant density. This is the purpose of the analysis of the PITs, introduced by Diebold, Gunther, and Tay~\cite{DGT} and widespread in the literature of evaluation of density forecasts~\cite{GBR,HO,KP}. We now expose this method, that we will adapt in the next subsection from the evaluation of density forecasts to the selection of the bandwidth and of the discount factor.

We observe $T$ successive price returns: $X_1,...,X_T$. We use the $t_0$ first to build a density estimation $\hat f^{h,\omega}_{t_0}$, using equation~\eqref{eq:InitDens}. This density includes a discount in order to depict more closely recent observations. We thus conceive $\hat f^{h,\omega}_{t_0}$ as a forecast of the true density $f_{t_0+1}$ of $X_{t_0+1}$, as well as we conceive $\hat f^{h,\omega}_t$, for any $t\geq t_0$, as a forecast of the true density $f_{t+1}$ of $X_{t+1}$. Of course, $f_t$ varies with $t$, and we only observe one random variable in this density, namely $X_t$. However, we are able to build a density which does not depend on $t$ and which will therefore be very useful for evaluating the quality of the density forecast. This invariant distribution is the one of the PIT variables, which are defined by:
$$Z^{h,\omega}_t=\hat F^{h,\omega}_{t-1}(X_t).$$
Indeed, if our forecast is good, that is if $\forall t\geq t_0+1$, $\hat F^{h,\omega}_{t-1}=F_t$, where $F_t$ is the true cdf, then, whatever $t$, $Z^{h,\omega}_t$ follows a uniform random variable in the interval $[0,1]$: $Z^{h,\omega}_t\sim\mathcal U(0,1)$~\cite{DGT}. In fact, this idea is quite old~\cite{Rosenblatt} and is even something with which any person simulating random variables following a given cdf is familiar. In addition to being uniform, the variables $Z^{h,\omega}_{t_0+1},...,Z^{h,\omega}_T$ must also be independent.

Thanks to the PITs, we have $T-t_0$ observations in the same uniform density. This makes it possible to evaluate the density forecast: we have to check that $Z^{h,\omega}_{t_0+1},...,Z^{h,\omega}_T$ are indeed iid and uniform in $[0,1]$. We expose in the next subsection the difference of approach regarding this point between the evaluation of density forecast and our framework, which is about the selection of optimal parameters.

\subsection{Selection of the bandwidth and of the discount factor}\label{sec:bandwidth}

The literature about bandwidth selection is very rich~\cite{WJ,JMS,Tsybakov}. Beyond the rule-of-thumb which often leads the choice of the bandwidth $h$ among practitioners, we can cite more relevant methods of selection, such as the minimization of AMISE~\cite{Silverman}, evaluated for instance with cross validation or a plug-in technique. As exposed in the previous subsection, we will try to select $h$ in order to make the distribution of the PITs close to a uniform distribution, and the uniform case is trivial in the AMISE approach and makes this method ineffective. We can also cite the possibility to estimate a time-varying bandwidth, like in the literature about online estimation of kernel density~\cite{WD,KLS}. Our approach will be different from the online framework: our time-varying aspect is not about $h$ but about the density. 

Our problem, in addition, is not only about selecting $h$. We have to select it jointly with the discount factor $\omega$. As already mentioned, we can base this selection on the maximization of a likelihood~\cite{HO}. But we want to have an accurate description of the true density, not to make the best point forecast. This thus incites us to use PITs and to adapt the method of evaluation of density forecasts. We have two objectives regarding the PITs: the uniformity and the independence.

We first focus on the uniformity. According to Diebold, Gunther, and Tay, methods based on statistical tests, such as the Kolmogorov-Smirnov test, are not relevant because nonconstructive, insofar as they do not indicate why PITs are not uniform~\cite{DGT}. They thus prefer a qualitative analysis using graphical tools such as a simple correlogram. Besides this mainstream approach, some papers propose a statistical test assessing the uniformity of the PITs~\cite{Berkowitz}. Our framework is in fact different as we do not want to determine whether our density forecast is good or not. Instead, given a density model, we only want to select its best parameters, here $h$ and $\omega$. Maybe our forecast will be poor, even though the non-parametric approach makes this case unlikely, but we will have done the best with respect to the model used. We thus do not want to test the consistence of our PITs with a uniform distribution, but we select the parameters $h$ and $\omega$ minimizing some test statistic. We choose to minimize the Kolmogorov-Smirnov statistic, $k$, because it is widespread and easy to understand. This statistic is simply the maximum gap between the empirical cdf and the theoretical one, which, in our case, is uniform:
\begin{equation}\label{eq:k}
k(Z^{h,\omega}_{t_0+1},...,Z^{h,\omega}_T)=\underset{t_0+1\leq s\leq T}{\max}\left|Z^{h,\omega}_{s}-\frac{1}{T-t_0+1}\sum_{u=t_0+1}^{T}{\indic_{[0,Z^{h,\omega}_s]}(Z^{h,\omega}_u)}\right|.
\end{equation}


But the Kolmogorov-Smirnov statistic says nothing about the independence of the variables and the fact that the sampling is random~\cite{DTW}. This property is however crucial and its absence could lead to nonsense estimations~\cite{HE,Davis}. In the standard approach regarding the evaluation of density forecast, the independence is assessed by graphical tools, such as a correlogram~\cite{DGT}. We would like again a more systematic approach. We thus use an additional criterion coming from the literature of simulation of quasi-random variables. We indeed want our series of PITs to be a low-discrepancy sequence~\cite{Niederreiter,Tuffin,Niederreiter2,GST}. This is even more important that we want to estimate a time-varying density of price returns in a regime-switching market. The rationale behind the discrepancy is that the uniformity must be a feature not only of isolated PITs but also of vectors of PITs: the sequence must be equidistributed. This method will avoid almost static densities in which price returns are globally well distributed, but with mainly high PITs in a bullish regime and then mainly low PITs during a crisis period. The discrepancy criterion we propose to minimize is then the multivariate uniformity Kolmogorov-Smirnov statistic for a given size of vectors of PITs, taking into account the targeted independence of the PITs. This independence, along with the uniformity, makes that the theoretical joint distribution of the vector $(Z^{h,\omega}_s,Z^{h,\omega}_t)$ is $Z^{h,\omega}_sZ^{h,\omega}_t$ in the targeted case, as soon as $s\neq t$. We focus on vectors of dimension 2 because the discrepancy statistics are difficult to compute for larger dimensions~\cite{JPZ,LFH}. In particular, for a pair of observations with a given time lag $\tau> 0$, we define:
$$k_{\tau}(Z^{h,\omega}_{t_0+1},...,Z^{h,\omega}_T)=\underset{t_0+1\leq s\leq T-\tau}{\max}\left|Z^{h,\omega}_{s}Z^{h,\omega}_{s+\tau}-\frac{1}{T-\tau-t_0+1}\sum_{u=t_0+1}^{T-\tau}{\indic_{[0,Z^{h,\omega}_s]}(Z^{h,\omega}_u)\indic_{[0,Z^{h,\omega}_{s+\tau}]}(Z^{h,\omega}_{u+\tau})}\right|.$$
In the definition of $k_{\tau}$, the case $\tau=0$ is not allowed. Indeed, in this case, the two PITs $Z^{h,\omega}_{s}$ and $Z^{h,\omega}_{s+\tau}$ are not to be independent from each other because they are equal. However, we can define another statistic $k'_{\tau}$, for $\tau\geq 0$, which includes both the independence of the PITs for $\tau>0$, thanks to $k_{\tau}$, and their univariate uniformity, thanks to $k$:
$$k'_{\tau}(Z^{h,\omega}_{t_0+1},...,Z^{h,\omega}_T)=\left\{\begin{array}{ll}
k(Z^{h,\omega}_{t_0+1},...,Z^{h,\omega}_T) & \text{if } \tau=0 \\
k_{\tau}(Z^{h,\omega}_{t_0+1},...,Z^{h,\omega}_T) & \text{else}.
\end{array}\right.$$
Besides, the multivariate Kolmogorov-Smirnov statistic depends on the size of the sample. We thus consider a size-adapted aggregation of the $(k'_{\tau})_{\tau\geq 0}$. Indeed, for a sample of $n\rightarrow +\infty$ observations, $\sqrt{n}\times k'_{\tau}$ has a limit distribution which does not depend on $n$~\cite{Deheuvels}. We also choose a maximal lag $\nu$ above which we will not consider dependence effects. Indeed, if the time lag is too big the number of observations is very limited and the asymptotic Kolmogorov distribution may not apply.\footnote{ We may consider, for example, $\nu=22$, so that we verify the independence for every one-month interval of every pair of daily price returns. This arbitrary choice is applied in the empirical part of this paper.} We thus propose the following size-adapted statistic:\footnote{ An alternative criterion is proposed in Appendix~\ref{sec:Alt}.}
$$d_{\nu}(Z^{h,\omega}_{t_0+1},...,Z^{h,\omega}_T)=\underset{0\leq\tau\leq\nu}{\max}\left(\sqrt{T-\tau-t_0}\times k'_{\tau}(Z^{h,\omega}_{t_0+1},...,Z^{h,\omega}_T)\right).$$

Finally, the optimal bandwidth and discount factor are defined as the parameters minimizing this uniformity and discrepancy statistic:
\begin{equation}\label{eq:optimal}
(h^{\star},\omega^{\star})=\underset{h>0, 0<\omega\leq 1}{\text{argmin}} d_{\nu}(Z^{h,\omega}_{t_0+1},...,Z^{h,\omega}_T).
\end{equation}
By doing so, we do not exactly target the independence of any vector of observations. Instead, our simplified statistic aims at obtaining the independence of pairs of observations contained in a time interval of duration $\nu$, like with the correlogram approach suggested by Diebold, Gunther, and Tay~\cite{DGT}.

We also propose a constrained version of this optimisation problem. Indeed, the above unconstrained problem may lead to a dynamic of densities far from the economic intuition, for example with very rough densities. In practice, the time-varying densities we have built with this method seem empirically robust, at first sight. But we are interested in defining some reasonable bounds for $h$ or $\omega$. In order to have a robust time-varying density, we want that an isolated observation does not change too much the density between two consecutive dates. To state things quantitatively, we want to limit the Kolmogorov-Smirnov statistic between two densities at consecutive dates. We propose $\nu^{-1}$ as an upper bound. With a daily $\nu^{-1}$ bound, the maximal change of the Kolomogorov-Smirnov statistic, which is 1, cannot be reached before the horizon $\nu$. The link between the $\nu^{-1}$ bound and the parameters is straightforward, as the update of the cdf leads to an increase of the cdf at one point of at most $1-\omega$. Therefore, we introduce a new bound for $\omega$ and the constrained problem is as follows:
\begin{equation}\label{eq:optimalConstr}
(h^{\star}_{c},\omega^{\star}_{c})=\underset{h>0, 1-\nu^{-1}<\omega\leq 1}{\text{argmin}} d_{\nu}(Z^{h,\omega}_{t_0+1},...,Z^{h,\omega}_T).
\end{equation}
We could also want to set bounds to $h$ in order to secure the robustness of the density at one date instead of the robustness across time. Nevertheless, we consider that the robustness across time is enough. Indeed, as the density will not change very rapidly, each density will be a fairly good forecast for observations close in time.

\subsection{Amplitude of the variations of the series of densities}\label{sec:divergence}

In the method exposed above to select $h$ and $\omega$, we minimize the divergence between an empirical distribution and a uniform one. In particular, we use the Kolmogorov-Smirnov statistic to depict this divergence because of both its simplicity and its asymptotic behaviour. But other divergence metrics could replace the Kolmogorov-Smirnov statistic in this method. 

We can also use various divergence statistics to quantify to which extent the estimated pdf is different from what it was at a reference date and thus track the evolution of the pdf through time. This is what these various statistics are devoted to in the empirical part of this paper. In addition, thanks to simulations, we will determine confidence intervals for each of these divergences at each date, so that we will be able to assess whether the evolution of the pdf through time is significant or not. We now review three of these divergence statistics in addition to the Kolmogorov-Smirnov statistic. Some are based on a comparison of densities, and others on a comparison of cdfs or even of quantiles.

First we recall the definition of the Kolmogorov-Smirnov statistic between the cdfs $\hat F_{t}$ and $\hat F_{t_0}$:
$$k\left(\hat F_{t},\hat F_{t_0}\right)=\underset{x\in\mathbb R}{\sup}\left|\hat F_{t}(x)-\hat F_{t_0}(x)\right|.$$

Whereas the Kolmogorov-Smirnov statistic considers the maximal difference between two cdfs, the Hellinger distance is the cumulated difference between densities:
$$H\left(\hat f_{t},\hat f_{t_0}\right)=\sqrt{\frac{1}{2}\int_{\mathbb R}{\left(\sqrt{\hat f_{t}(x)}-\sqrt{\hat f_{t_0}(x)}\right)^2dx}}.$$

The $p$-Wasserstein distance, given $p\geq 1$, is related to the optimal transportation theory~\cite{Villani}. It is the minimal cost to reconfigure one pdf in another one. The Kolmogorov-Smirnov statistic is the $L^{\infty}$ distance between the cdfs, whereas the Wasserstein metric is the $L^{p}$ distance between their quantiles. The $p$-Wasserstein distance is indeed defined by~\cite{PZ}:
$$W_p\left(\hat F_{t},\hat F_{t_0}\right) = \left(\int_0^1{\left|\hat F_{t}^{-1}(\alpha)-\hat F_{t_0}^{-1}(\alpha)\right|^p d\alpha}\right)^{1/p}.$$
In this paper, we focus on the case $p=1$, for which the Wasserstein distance is also equal to the $L^{1}$ distance between cdfs~\cite{PZ}. It thus clearly generalizes the Kolmogorov-Smirnov statistic. We will see in the empirical part of this paper that this generalization may be more appropriate than the Kolmogorov-Smirnov statistic to assess the significance of the variations of the distribution. Indeed, the occurrence of several extreme observations may not impact significantly the Kolmogorov-Smirnov statistic, which mainly focuses on the body of the distribution, whereas the Wasserstein distance takes into account the whole distribution. On the other hand, since a uniform distribution is not subject to a dichotomy between body and tails, the Kolmogorov-Smirnov statistic seems appropriate for assessing the uniformity of the PITs.

As opposed to the other divergences, the Kullback-Leibler divergence is not strictly speaking a distance function as it is not symmetric in the two densities. It is related to Shannon's entropy. It is defined by:
$$D_{KL}\left(\hat f_{t}\right|\left|\hat f_{t_0}\right) = \int_{\mathbb R}{\hat f_{t}(x)\log\left(\frac{\hat f_{t}(x)}{\hat f_{t_0}(x)}\right)dx}.$$ 

All these divergences can be generalized easily if we work with a discrete grid instead of $\mathbb R$. They are also always positive and with a value of zero if $\hat f_{t}=\hat f_{t_0}$.

\section{A simulation study}\label{sec:simul}

The purpose of this simulation study is to confirm with simulated data the relevance of the selection method for $h$ and $\omega$, introduced in equation~\eqref{eq:optimal}. We thus introduce an alternative selection criterion and compare the performance of the two methods in forecasting a simple simulated time-varying density. 

The alternative selection method we use, in this time-varying framework, is the one proposed by Harvey and Oryshchenko. They use a maximum likelihood approach to select the free parameters $h$ and $\omega$~\cite{HO}:
\begin{equation}\label{eq:critHO}
(h^{HO},\omega^{HO})=\underset{h>0, 0<\omega\leq 1}{\text{argmax}} \sum_{s=t_0}^{T-1}{\log\left(\hat f^{h,\omega}_{s}(X_{s+1})\right)}.
\end{equation}

We now compare the two methods defined by equations~\eqref{eq:optimal} and~\eqref{eq:critHO}. It appeared to us that our approach performs well when the simulated distribution has fat tails. Therefore, we generate $2,000$ independent variables of time-varying density $f_t$, which is a Cauchy density with scale parameter equal to 1 and time-varying location parameter equal to $t/100$:
\begin{equation}\label{eq:cauchy}
f_t(x)=\frac{1}{\pi\left[1+\left(x-\frac{t}{100}\right)^2\right]}.
\end{equation}
Incidentally the $t$-th variable equivalently follows a Student's distribution with one degree of freedom, translated by a quantity $t/100$. 

The particularity of this kind of distribution is the frequent occurrence of high values, possibly very far from all the past observations. Likelihood-based selection methods of the free parameters of the kernel density thus require a consequent smoothing, that is a high $h$ and a high $\omega$, in order to avoid very low likelihoods for these big observations. On the contrary, a method based on a cdf criterion, such as the one we put forward, does not have a similar constraint. As a consequence, using the $t$ first simulated data to estimate the density at time $t+1$, for $t$ evolving between $1,000$ and $1,999$, we find smaller free parameters in the PIT-based approach than in the likelihood method: $h^{\star}=0.488$, $\omega^{\star}=0.902$, $h^{HO}=0.596$, and $\omega^{HO}=0.989$.

\begin{figure}[htbp]
	\centering
		\includegraphics[width=0.6\textwidth]{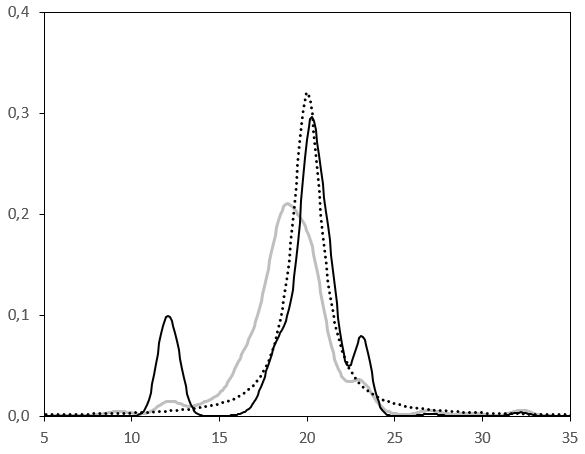} 
\begin{minipage}{0.7\textwidth}\caption{For the time-varying Cauchy distribution, true pdf (dotted line) and estimates at time $t+1=2,000$. The two solid lines represent the estimates with the two competing vectors of parameters: $(h^{\star},\omega^{\star})$ (black curve) and $(h^{HO},\omega^{HO})$ (grey).}
	\label{fig:cauchy_dens}
\end{minipage}
\end{figure}

\begin{figure}[htbp]
	\centering
		\includegraphics[width=0.45\textwidth]{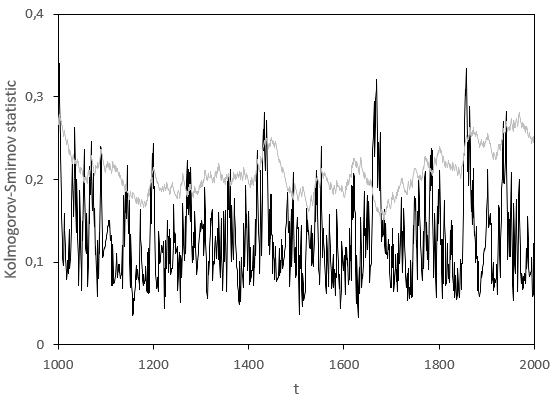} 
		\includegraphics[width=0.45\textwidth]{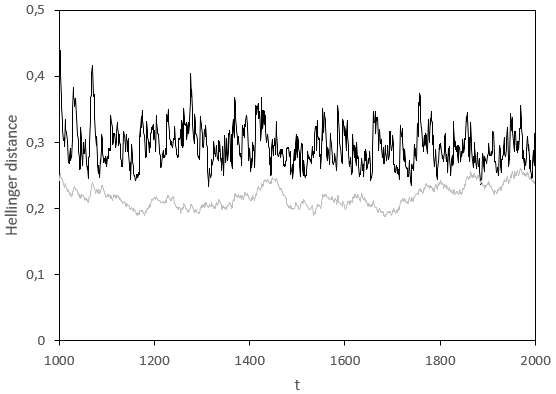} \\
		\includegraphics[width=0.45\textwidth]{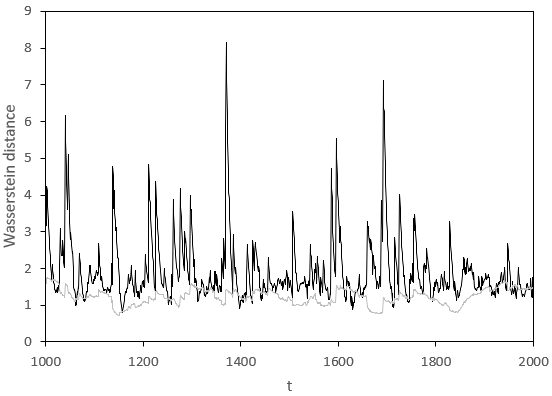} 
		\includegraphics[width=0.45\textwidth]{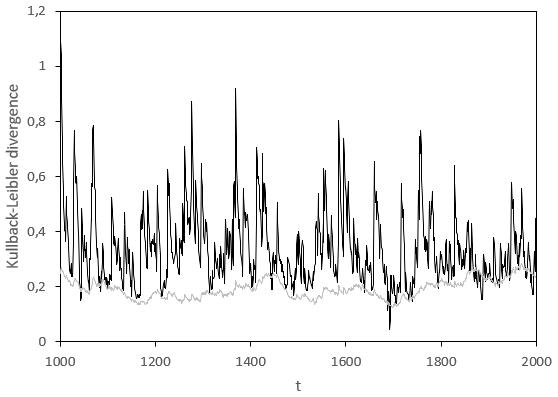} 
\begin{minipage}{0.7\textwidth}\caption{Divergence statistics with respect to $f_t$, as a function of the instant $t$, when $f_t$ is the time-varying Cauchy density defined in equation~\eqref{eq:cauchy}: the Kolmogorov-Smirnov statistic (top left), the Hellinger distance (top right), the Wasserstein distance (bottom left), and the Kullback-Leibler divergence (bottom right). The two curves represent the two competing vectors of parameters: $(h^{\star},\omega^{\star})$ (black curve) and $(h^{HO},\omega^{HO})$ (grey). The more erratic aspect of the black curves is to be explained by the lower selected free parameters.}
	\label{fig:cauchy_div}
\end{minipage}
\end{figure}

Since we know the true density, we can compare the estimated densities in $t$ with $f_{t+1}$. We see for example in Figure~\ref{fig:cauchy_dens} that the likelihood method leads to a smoother density whose bulk is also more shifted, because of the time-varying aspect, than the PIT-based approach. The divergence statistics introduced in Section~\ref{sec:divergence} reveal that our method, which explicitly uses a Kolmogorov-Smirnov criterion, leads to a better estimate of the time-varying density than the likelihood method, but only for the Kolmogorov-Smirnov statistic, as one can see in Figure~\ref{fig:cauchy_div}. This suggests that our PIT-based approach could benefit from an adaptation to other metrics than the sole Kolmogorov-Smirnov criterion, depending on the preferred divergence statistic. This could be the subject of further research.

One could wonder if the relative superiority, at least regarding the Kolmogorov-Smirnov statistic, of the PIT-based approach over the likelihood method also holds when the true density is static. We thus simulate $2,000$ i.i.d. Cauchy variables, with static density equal to $f_0$. We estimate a static kernel density $\hat f^h$ and cdf $\hat F^h$, like in equation~\eqref{eq:KernelStatique}, using only the $1,000$ first observations. This estimate is based on a bandwidth $h$ that we select following either a PIT validation criterion,
$$h^{\star,\text{static}}=\underset{h>0}{\text{argmin}}\ d_{\nu}(\hat F^{h}(X_{t_0+1}),...,\hat F^{h}(X_{T})),$$
or a likelihood criterion,
$$h^{HO,\text{static}}=\underset{h>0}{\text{argmax}} \sum_{s=t_0}^{T-1}{\log\left(\hat f^{h}(X_{s+1})\right)},$$
evaluated on the $1,000$ last observations of our simulated dataset. In other words, we use a straightforward adaptation of equations \eqref{eq:optimal} and~\eqref{eq:critHO} in the static framework.

Like in the dynamic framework, the PIT-based method leads to a less smooth density than the likelihood method: $h^{\star,\text{static}}=0.287$ and $h^{HO,\text{static}}=1.449$. All the divergence statistics with respect to the true density underline the superiority, in this case, of the PIT-based approach: 0.027 (0.093 for the likelihood method) for the Kolmogorov-Smirnov statistics, 0.113 (0.147) for the Hellinger distance, 0.513 (0.890) for the Wasserstein distance, and 0.032 (0.076) for the Kullback-Leibler divergence.

\section{Empirical results}\label{sec:EmpRes}

We now apply the PIT-based method to the estimation of time-varying densities of several stock indices. We consider American indices (NASDAQ Composite, S\&P 500, S\&P 100), European indices (EURO STOXX 50, Euronext 100, DAX, CAC 40), and Asian indices (Nikkei 225, KOSPI, SSE 50), with a particular focus on S\&P 500, EURO STOXX 50, and the South-Korean KOSPI indices. We have used data from Yahoo finance in the time interval from 04/17/2015 to 05/28/2020. The study period includes the economic crisis related to the COVID-19. In particular, we study the impact of the COVID-19 on three stock markets corresponding to economic areas with different crisis management regarding the pandemic. The questions we want to answer are about the significance of this impact and the characterization of a recovery after the peak of the crisis.

We have estimated daily a pdf of daily price returns from the date $t_0$ corresponding to November 1st, 2019. These densities include observations from 2015, exponentially weighted with an optimal discount factor depending on the index. We provide these optimal discount factors in Table~\ref{tab:optimal}, along with the optimal bandwidth, determined by equation~\eqref{eq:optimal}, as well as the constrained version defined by equation~\eqref{eq:optimalConstr}.

\begin{table}[htbp]
\centering
\begin{tabular}{l||c|c||c|c}
  & $h^{\star}$ & $\omega^{\star}$ & $h_c^{\star}$ & $\omega_c^{\star}$ \\
\hline
\hline
NASDAQ Composite & $6.5\times 10^{-3}$ & 0.875 & $3.1\times 10^{-4}$ & 0.962 \\
S\&P 500 & $1.0\times 10^{-5}$ & 0.864 & $6.9\times 10^{-3}$ & 0.955 \\
S\&P 100 & $4.3\times 10^{-3}$  & 0.889 & $2.8\times 10^{-3}$  & 0.963 \\
\hline
EURO STOXX 50 & $6.9\times 10^{-3}$ & 0.883 & $1.2\times 10^{-2}$ & 0.964 \\
Euronext 100 & $8.2\times 10^{-3}$ & 0.872 & $6.5\times 10^{-3}$ & 0.956 \\
DAX & $4.4\times 10^{-3}$ & 0.856 & $1.0\times 10^{-2}$ & 0.959 \\
CAC 40 & $1.1\times 10^{-2}$ & 0.780 & $8.7\times 10^{-3}$ & 0.957 \\
\hline
Nikkei 225 & $8.2\times 10^{-3}$ & 0.911 & $1.1\times 10^{-2}$ & 0.965 \\
KOSPI & $4.5\times 10^{-3}$ & 0.884 & $7.3\times 10^{-3}$ & 0.957 \\
SSE 50 & $4.1\times 10^{-3}$ & 0.914 & $8.2\times 10^{-3}$ & 0.974 
\end{tabular}
\begin{minipage}{0.7\textwidth}\caption{Optimal bandwidth $h^{\star}$ and discount factor $\omega^{\star}$ minimizing the criterion $d_{\nu}$, for several stock indices for densities between November 2019 and May 2020. The constrained version is $h_c^{\star}$ and $\omega_c^{\star}$.}
\label{tab:optimal}
\end{minipage}
\end{table}


For the rest of the empirical study, we consider a common pair of parameters for all the indices, so as to make fair comparisons. Focusing on the constrained case reported in Table~\ref{tab:optimal}, we choose the highest estimated bandwidth, to ensure robustness of the densities, and the lowest discount parameter, so as to have the highest responsiveness of the dynamics of densities: $h^m=0.012$ and $\omega^m=0.955$. We remark that these values are close to the median values obtained for an alternative method described in Appendix~\ref{sec:Alt}.

\begin{figure}[htbp]
	\centering
		\includegraphics[width=0.45\textwidth]{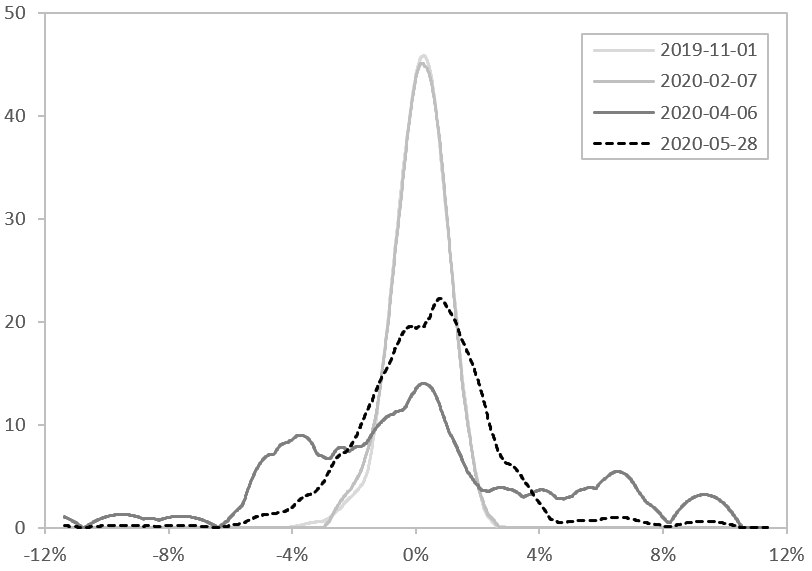} 
		\includegraphics[width=0.45\textwidth]{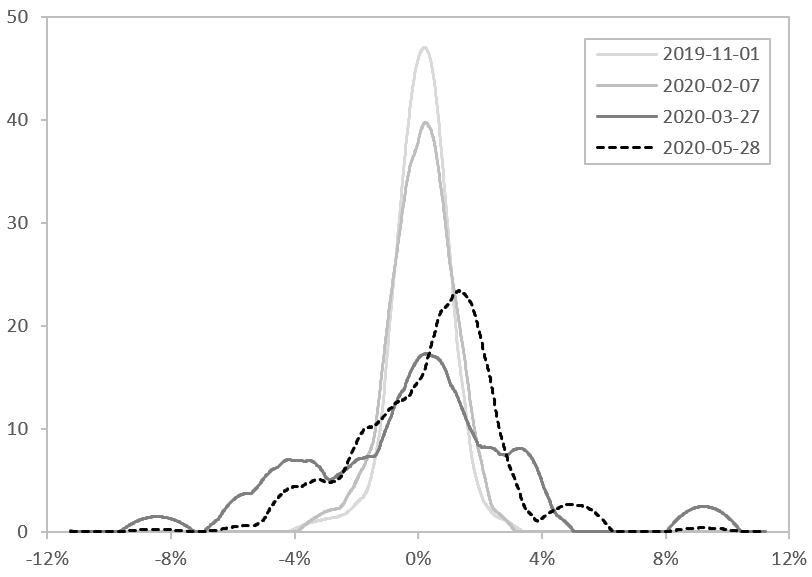} \\
		\includegraphics[width=0.45\textwidth]{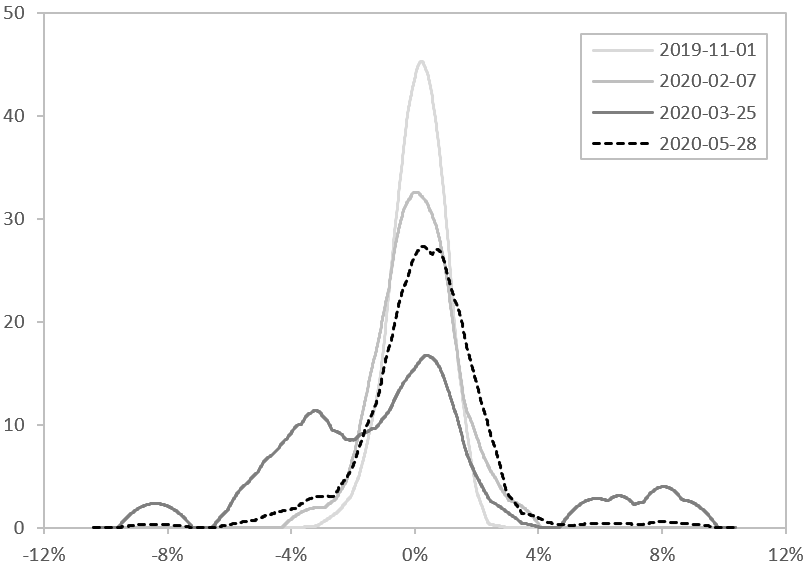} 
\begin{minipage}{0.7\textwidth}\caption{Estimated dynamic pdf of daily price returns for S\&P 500 (top left), EURO STOXX 50 (top right), and KOSPI (bottom) indices.}
	\label{fig:pdf}
\end{minipage}
\end{figure}

For the S\&P 500, EURO STOXX 50, and KOSPI indices, we display in Figure~\ref{fig:pdf} the estimated dynamic pdf of price returns at four dates which illustrate the chronology of the impact of the pandemic on financial markets:
\begin{itemize}
\item before the crisis: on the 16th December 2019, in a period where the markets were steady,
\item at the first turmoil in the markets, the 7th February 2020,
\item at the peak of the pandemic, which occurs at a different date for each market,
\item at the end of our sample, the 28th May 2020.
\end{itemize}
We determine the date of the peak of the pandemic as the date $t$ maximizing the Hellinger distance of the density $\hat f^{h^m,\omega^m}_t$ with respect to the estimated pdf in $t_0$, that is $\hat f^{h^m,\omega^m}_{t_0}$. This peak does not follow an epidemiological definition, since we only observe financial data. It corresponds to a maximal divergence with the steady state of the market.

For the KOSPI and the EURO STOXX 50, the pdf before the crisis looks like a Gaussian distribution, with thin tails. Then the pdf slightly widens on the losses side. At the peak of the crisis, the pdf crushes, with very fat tails. After the peak, it tends to an asymmetric distribution with a negative skewness and slowly decreasing tails. The chronology is similar for the S\&P500, except that the pdf the 7th February is similar to the one before the crisis. It may indicate a low responsiveness of the US market in front of the outbreak. Or it may denote a temporary lag in the impact of the COVID-19 on the US market, reflecting the lag in the spread of the outbreak in the region.

\begin{figure}[htbp]
	\centering
		\includegraphics[width=0.45\textwidth]{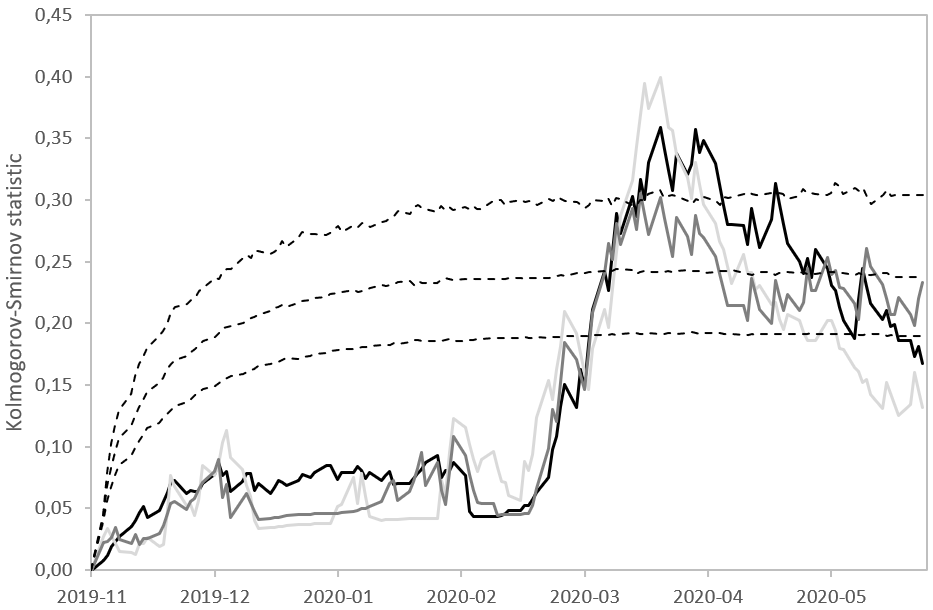} 
		\includegraphics[width=0.45\textwidth]{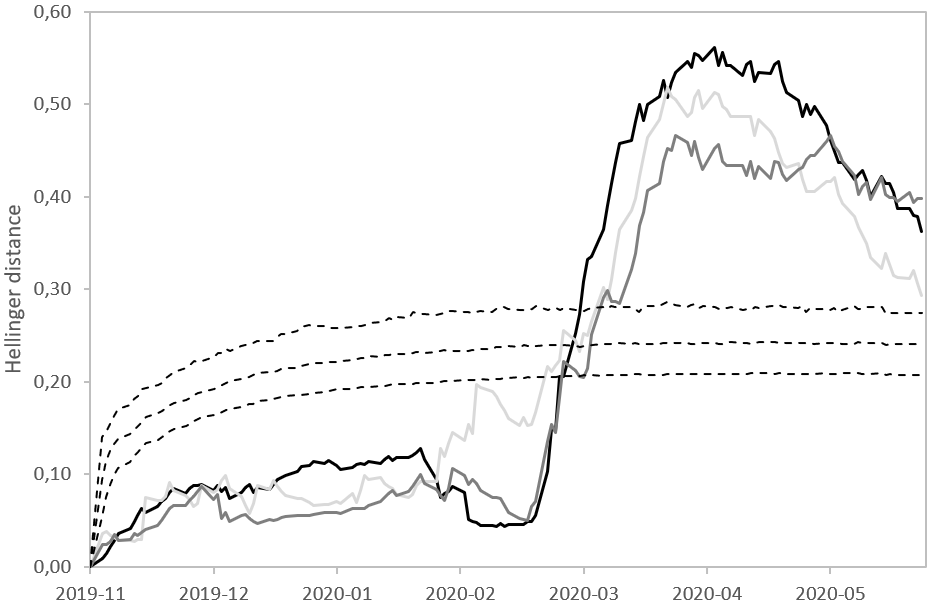} \\
		\includegraphics[width=0.45\textwidth]{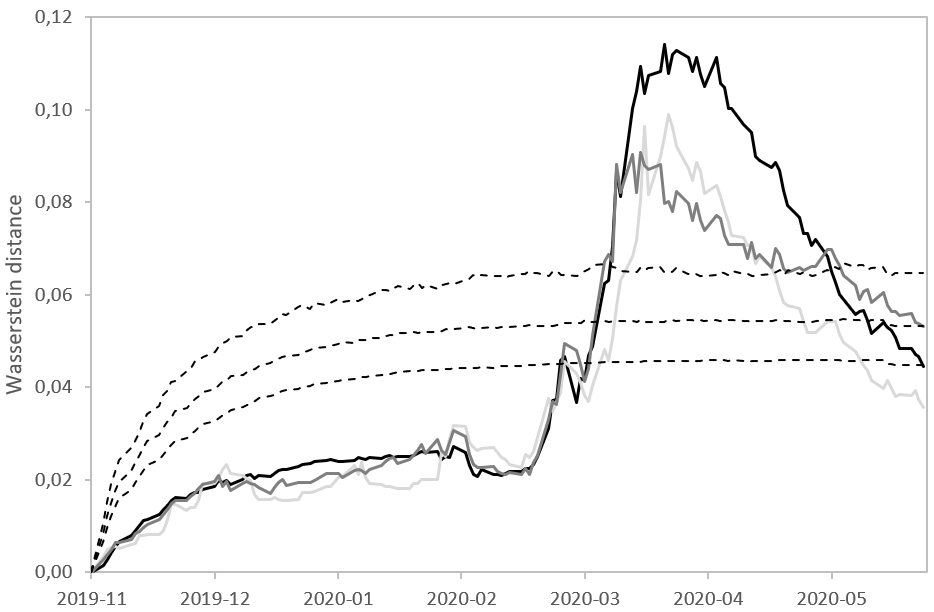}  
		\includegraphics[width=0.45\textwidth]{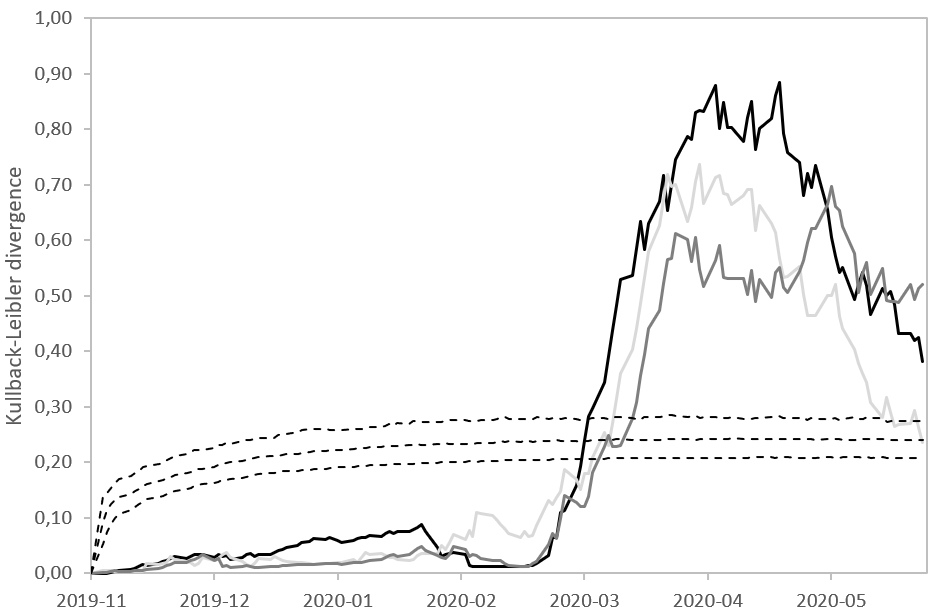}  
\begin{minipage}{0.7\textwidth}\caption{Daily evolution through time of four divergence statistics: the Kolmogorov-Smirnov statistic (top left), the Hellinger distance (top right), the Wasserstein distance (bottom left), and the Kullback-Leibler divergence (bottom right). The curves correspond to S\&P 500 (black), EURO STOXX 50 (dark grey), and KOSPI (light grey) indices. The dotted lines are simulated confidence intervals, with confidence levels, from the bottom to the top: $95\%$, $99\%$ and $99.9\%$.}
	\label{fig:divergence}
\end{minipage}
\end{figure}

Displaying pdfs at several dates as in Figure~\ref{fig:pdf} makes it possible to depict the chronology of the crisis. But it is limited since displaying this density every day of our sample would make the figure unreadable. Therefore, instead of displaying each pdf, we display one statistic per day. This statistic must reflect the divergence of the pdf with respect to a steady state of markets. We thus determine the Kolmogorov-Smirnov statistic, the Hellinger distance, the $1$-Wasserstein distance, as well as the Kullback-Leibler divergence of the pdf each day with respect to the pdf in $t_0$. Results are displayed in Figure~\ref{fig:divergence}. Whatever the divergence statistic, we observe first a slight increase from 0 toward a low positive value until the beginning of the crisis, where the divergence sharply increases till the peak where it begins to slowly decrease. This last phase corresponds to the slow recovery of the markets after the crisis. 

In addition to the evolution through time of the four divergence statistics, Figure~\ref{fig:divergence} shows confidence intervals for each statistic. These confidence intervals come from the simulation of 10,000 Brownian motions on which we apply our method of density estimation and implementation of divergence statistics. The null hypothesis $H_0$ is thus that all the price returns are iid Gaussian variables. For each statistic and each date, we represent the quantile estimated on simulations and corresponding to three confidence levels: $95\%$, $99\%$ and $99.9\%$. At a given date $t$, for a given stock index, if the divergence of the current pdf with respect to the pdf in $t_0$ is above a particular curve of the confidence interval, we reject $H_0$ with the corresponding confidence level $p$. In other words, we consider the pdf in $t$ to be significantly different from the pdf in $t_0$ with a confidence level $p$.

Depending on the divergence considered, we are able to determine the peak of the impact as the date maximizing the statistic. We display in Table~\ref{tab:Hellinger} the date of the peak as well as the value of the divergence at the peak, before the crisis, and late May in the Hellinger approach. According to this table, the strongest impact is in the US but the recovery seems faster there than in Europe. The smallest impact and the fastest recovery is by far in China. The peak occurs between the 25th March (China and South-Korea) and the 6th April (US and Japan), whatever the index, except for the DAX, whose peak is in May. We use the other divergences as a robustness check of these results. The conclusions are in fact similar: small impact and almost total recovery late May for the Chinese SSE 50 index, strongest impact on the US market, slowest recovery on the European market. We also observe some variations in the estimation of the peak date. The most surprising one is provided by the Kolmogorov-Smirnov statistic, according to which the peak is reached first in Europe the 18th March, before continental Asia and US the 23rd March, and finally Japan the 2nd April. 

\begin{table}[htbp]
\centering
\begin{tabular}{l||c|c|c|c}
  & $H$ on 7th Feb. & $H$ at the peak & Date of the peak & $H$ on 28th May \\
\hline
\hline
NASDAQ Composite & 0.111 & 0.531 & 2020-04-01 & 0.295 \\
S\&P 500 & 0.084 & 0.562 & 2020-04-06 & 0.363 \\
S\&P 100 & 0.089 & 0.562 & 2020-04-06 & 0.324 \\
\hline
EURO STOXX 50 & 0.051 & 0.466 & 2020-03-27 & 0.398 \\
Euronext 100 & 0.050 & 0.484 & 2020-03-27 & 0.385 \\
DAX & 0.061 & 0.458 & 2020-05-05 & 0.377 \\
CAC 40 & 0.052 & 0.479 & 2020-03-27 & 0.389 \\
\hline
Nikkei 225 & 0.095 & 0.477 & 2020-04-06 & 0.350 \\
KOSPI & 0.083 & 0.518 & 2020-03-25 & 0.294 \\
SSE 50 & 0.070 & 0.381 & 2020-03-25 & 0.122 
\end{tabular}
\begin{minipage}{0.7\textwidth}\caption{Hellinger distance $H$ with respect to $t_0$. The peak corresponds to when the maximal Hellinger distance is reached. Dates are in 2020.}
\label{tab:Hellinger}
\end{minipage}
\end{table}

We stress the fact that the alternative chronology of the peaks is not the only particularity of the Kolmogorov-Smirnov statistic with respect to the three other divergence statistics we have implemented. For instance, the significance of the financial crisis in some regions is questionable according to this divergence statistic, as one can see in Figure~\ref{fig:divergence}. We can nevertheless explain this striking, and certainly dubious, conclusion. Indeed, when simulating two sets of iid random variables, we get two kernel densities but the Kolmogorov-Smirnov statistic focuses on only one quantile, generally corresponding to where the cdf is the steepest, that is in the body of the distribution. If we disrupt one of these two densities with a limited number of outliers, the Kolmogorov-Smirnov statistic may not change a lot as this modification mainly impacts the tails and not the body of the distribution. On the contrary, the three other divergence statistics are less robust to outliers as they are defined by integrals over all the distribution. Their responsiveness to a crisis is thus higher. For this reason, we prefer them to the Kolmogorov-Smirnov statistic for assessing the significance of the variations of a dynamic pdf.

\section{Conclusion}\label{sec:concl}

In this paper, we have introduced a new method to select the two free parameters of a dynamic kernel density estimation, namely the discount factor and the bandwidth. This method relies on the maximisation of the accuracy of the daily pdf. This accuracy is to be understood in the sense of the literature about density forecast evaluation: the PIT of each new observation, expressed using the time-varying distribution, forms a set of variables which must be iid uniform variables. We use the Kolmogorov-Smirnov statistic and a discrepancy statistic to build a quantitative criterion of accuracy of the pdf. It is this criterion that we try to maximize when selecting the bandwidth and the discount factor of our time-varying pdf. Future research could focus on extensions of this method to other divergence statistics.

We have applied this method to financial data. In particular, we represent the evolution of the pdf of daily price returns for several stock indices during the COVID-19 pandemic. We are thus able to expose an accurate chronology of the financial crisis. Though the impact of the pandemic on the Chinese market seems limited, we observe that the strongest impact occurred in the US. The slowest recovery is in Europe, for which the pdf of daily returns is still significantly different from a steady market late May 2020. On the contrary, the recovery of the Chinese and South-Korean markets is very rapid. According to several divergence statistics late May 2020, they are even not significantly different from what they were before the crisis.

\bibliographystyle{plain}

\bibliography{biblioDynDens}

\appendix

\section{An alternative criterion}\label{sec:Alt}

The selection of the free parameters $h$ and $\omega$ relies on the minimization of a criterion $d_{\nu}$ exposed in Section~\ref{sec:bandwidth}. Alternatively to this criterion, we propose another criterion $\mathcal{D}_{\nu}$ to be minimized. The difference between $d_{\nu}$ and $\mathcal{D}_{\nu}$ consists in a different interpretation of the independence of the PITs. In $\mathcal{D}_{\nu}$, one simply considers a necessary condition of the independence, what makes this approach less rigorous than $d_{\nu}$. Given the statistic $k$ defined in equation~\eqref{eq:k}, the criterion $\mathcal{D}_{\nu}$ follows the definition:
$$\mathcal D_{\nu}(Z^{h,\omega}_{t_0+1},...,Z^{h,\omega}_T)=\underset{t_0+1\leq s<s+\nu-1\leq t\leq T}{\max}\left(\sqrt{t-s+1}\times k(Z^{h,\omega}_s,...,Z^{h,\omega}_t)\right).$$
The rationale behind this alternative criterion is that the uniformity must be a feature not only of the PITs in the interval $[t_0+1,T]$ but also of the PITs in any of its subintervals: the sequence must be equidistributed. Like for $d_{\nu}$, this method will avoid almost static densities with price returns globally well distributed, but with high PITs in a bullish regime compensated by low PITs during a crisis period. The criterion $\mathcal D_{\nu}$ to be minimized is then the worst uniformity statistic over all the subintervals of $[t_0+1,T]$. Once again, this criterion adapts the Kolmogorov-Smirnov statistics to the size of the subsample~\cite{MTW}. We also choose a minimal size $\nu$ above which we consider that the asymptotic Kolmogorov distribution may be applied. We may consider, for example, $\nu=22$, so that we verify the uniformity for every one-month interval of daily price returns. Thanks to this size-adapted statistic, the criterion $\mathcal{D}_{\nu}$ in fact focuses on the subinterval of size higher than $\nu$ with the least uniform PITs.

The optimal parameters $h$ and $\omega$ obtained by this method for the stock indices studied in the empirical part of this paper are gathered in Table~\ref{tab:optimalAlter}.

\begin{table}[htbp]
\centering
\begin{tabular}{l||c|c||c|c}
  & $h^{\star}$ & $\omega^{\star}$ & $h_c^{\star}$ & $\omega_c^{\star}$ \\
\hline
\hline
NASDAQ Composite & 0.0110 & 0.827 & 0.0101 & 0.955 \\
S\&P 500 & 0.0122 & 0.840 & 0.0121 & 0.955 \\
S\&P 100 & 0.0124  & 0.877 & 0.0124  & 0.955 \\
\hline
EURO STOXX 50 & 0.0124 & 0.831 & 0.0124 & 0.960 \\
Euronext 100 & 0.0110 & 0.883 & 0.0123 & 0.963 \\
DAX & 0.0038 & 0.864 & 0.0014 & 0.955 \\
CAC 40 & 0.0117 & 0.864 & 0.0119 & 0.960 \\
\hline
Nikkei 225 & 0.0124 & 0.790 & 0.0165 & 0.955 \\
KOSPI & 0.0124 & 0.813 & 0.0286 & 0.955 \\
SSE 50 & 0.0107 & 0.778 & 0.0002 & 0.974 \\
\hline
\hline
Mean value & 0.0110 & 0.838 & 0.0112 & 0.959 \\
Median value & 0.0122 & 0.840 & 0.0123 & 0.955
\end{tabular}
\begin{minipage}{0.7\textwidth}\caption{Optimal bandwidth $h^{\star}$ and discount factor $\omega^{\star}$ minimizing the criterion $\mathcal D_{\nu}$ for several stock indices for densities between November 2019 and May 2020. The constrained version is $h_c^{\star}$ and $\omega_c^{\star}$.}
\label{tab:optimalAlter}
\end{minipage}
\end{table}

\end{document}